\documentstyle[sprocl]{article}

\def\Journal#1#2#3#4{{#1} {\bf #2}, #3 (#4)}
\def\AP{{\em Ann. Phys.}}

\def\JPA{{\em J. of Phys.} A}

\def\NP{{\em Nucl. Phys.}}

\def\PRD{{\em Phys. Rev.} D}

\def\PRL{{\em Phys. Rev. Lett.}}
\def\PTP{{\em Progr. of Th. Phys.}}

\newcommand{\be}{\begin{equation}}
\newcommand{\ee}{\end{equation}}
\newcommand{\bea}{\begin{eqnarray}}
\newcommand{\eea}{\end{eqnarray}}

\newcommand{\hf} {{1\over2}}
\newcommand{\br} {\hskip -0.2cm /}
\newcommand{\bre} {\hskip -0.15cm /}
\newcommand{\nonu}{\nonumber\\}

\def\dt{\Delta\tau}
\def\ca{{\cal A}}
\def\eq#1{(\ref{#1})}

\begin{document}

\title{PATH INTEGRAL FOR THE DIRAC EQUATION}

\author{J. POLONYI}

\address{Laboratory of Theoretical Physics, Louis Pasteur University\\
3 rue de l'Universit\'e 67087 Strasbourg, Cedex, France\\
and\\
Department of Atomic Physics, L. E\"otv\"os University\\
Puskin u. 5-7 1088 Budapest, Hungary\\
E-mail: polonyi@fresnel.u-strasbg.fr}

\maketitle

\abstracts{A path integral representation is given
for the solutions of the 3+1 dimensional Dirac equation.
The regularity of the trajectories, the non-relativistic limit and 
the semiclassical approximation are briefly mentioned.}

There are different integral or sum representations for the
solution of the Dirac equation. The summation over trajectories
on 1+1 dimensional lattice \cite{sum} can be generalized by the help of the Grassman
variables \cite{grass} for 3+1 dimensions \cite{sch}. The method presented here
\cite{pierre} has the advantage that the integration is over the real 
trajectories in the continuous three-space, just as in the case of the 
non-relativistic Schr\"odinger equation. The disadvantage
is that the spin and the chirality flips are still described
by spinors so the integrand of the path integral consists of the
product of $4\times4$ matrices. But this complication
can be avoided and the c-number integration recovered
by means of additional compact variables \cite{progr}. 

We start with massless chiral fermions whose equation of motion is
$(\partial_0+c\partial\cdot\sigma)\phi=0$.
The $\hbar$-independence reflects that the causality is maintained
even on the level of the quantum fluctuations. 
We introduce the infinitesimal
time propagator $G(x,y;\dt)=e^{i\ca(x,y;\dt)}$
defined by
\be
\phi(x,\tau+\dt)=\int dyG(x,y;\dt)\phi(y,\tau),\label{infev}
\ee
where $\tau=tc$.
The choice of the action $\ca(x,y;\dt)$ can be motivated
in the following manner: (i) We expect a self reproducible propagator,
\be
e^{i\ca(x,y;\tau_1+\tau_2)}=\int dz
e^{i\ca(x,z;\tau_1)}e^{i\ca(z,y;\tau_2)},\label{gblock}
\ee
what suggests a quadratic dependence of $\ca$ in the coordinates.
(ii) Translational invariance further restricts the propagator to
\be
\ca(x,y;\dt)=\hf(x-y)^jA_{jk}(\dt)(x-y)^k+(x-y)^jB_j(\dt).
\ee
(iii) Rotational symmetry imposes $A_{jk}=A\delta_{j,k}$ and 
that the vector $B_j$ should be parallel to the spin,
$B_j\approx\sigma_j$. (iv) We want to keep the massless propagator
$\hbar$-independent so the only choice is $A=a/\dt^2$ and
$B_j=b\sigma_j/\dt$, where $a$ and $b$ are dimensionless constants.
The substitution of this propagator into \eq{infev} shows that the desired
equation of motion is recovered in the continuum limit, $\dt\to0$
for $a=b$. The chiral amplitude $\phi(x)$ is thus evolved
by the path integral
\bea
G(x,y;\tau)&=&\prod_j\left\{{\cal N}^{-1}\int dx_je^{
i\kappa\left[{1\over2}\left({x-y\over\dt}\right)^2
-{x-y\over\dt}\cdot\sigma\right]}\right\}\nonu
&=&\int D[x(t)]T\left\{e^{
i{\kappa\over\dt}\int d\tau\left[\hf\left({dx\over d\tau}\right)^2
-{dx\over d\tau}\cdot\sigma\right]}\right\},\label{pathiml}
\eea
where we have taken $a=b=\kappa$.
Notice the explicit linear divergence, $\dt^{-1}$, in the action,
in analogy with Quantum Field Theory where the bare, regulated action
contains the divergent coupling constants. 

The finite time propagator in the Fourier and real spaces is
\bea
\tilde G(p,\tau)&=&e^{-i{3\tau\over2\dt}}
e^{-i\tau\left(\dt{1\over2\kappa\hbar^2}p^2
+{1\over\hbar}p\cdot\sigma\right)},\nonu
G(x,y;\tau)&=&{\cal N}^{-1}
e^{i\kappa{\tau\over\dt}\left(\hf\left({x-y\over\dt}\right)^2
-{x-y\over\dt}\cdot\sigma\right)}.\label{kernelft}
\eea
The correct spectrum $E(p)=\pm|p|$ is recovered in the continuum limit
according to $\tilde G(p,\tau)$. $G(x,y;\tau)$ shows that whenever the
particle wanders off the light cone the diverging phase
of the exponent cancels the non-causal amplitude.

The equation of motion for massive fermions in the presence 
of an external potential $A_\mu(x)$,
$i\partial_\tau\psi=[\alpha\cdot(-i\partial+e/c\hbar A)
+\beta/\lambda-e/c\hbar A_0]\psi$
where $\lambda=\hbar/mc$, is generated by the path integral
\be
G(x,y;\tau)=\int D[x(t)]T\left\{e^{i{\kappa\over\dt}\int d\tau
\left[\hf\left({dx\over d\tau}\right)^2-{dx\over d\tau}\cdot\alpha\right]}
e^{i\int d\tau\left[{e\over\hbar c}\gamma^0A\bre-\beta/\lambda\right]}\right\},
\label{piaext}
\ee
with $A\br=\gamma^0A_0+\gamma\cdot A$.
One can show that the resulting propagator can be written for $A_\mu=0$ as
\be
G(x,y;\tau)=\int{dp\over(2\pi\hbar)^3}U(p)
e^{{i\over\hbar}p\cdot(x-y)-{i\over\hbar}\beta tmc^2
\sqrt{1+\left({p\over mc}\right)^2}}U^\dagger(p),
\ee
where $U(p)$ describes a basis transformation, so
it reproduces the usual relativistic spectrum.

The non-relativistic limit appears as a crossover between
the scaling laws 
\be
<\left({\Delta x\over\dt}\right)^2>=\cases{1&$\dt<<\lambda$,\cr
\lambda/\dt&$\dt>>\lambda$}
\ee
at $\dt\approx\lambda$. This result can be made more
plausible my noting that that the only dimensional
parameter for the average velocity in the relativistic 
region where mass is negligible is $c$.
Thus the trajectories of our path integral become
smoother then their non-relativistic counterpart, 
suggesting that the fractal-like propagation of
the first quantized non-relativistic quantum mechanics can be embedded into
a smoother, more "classical" relativistic quantum dynamics.

In order to obtain the saddle point approximation consider the amplitude
\be
\int D[x(t)]\eta^\dagger_fT\left\{e^{i{\kappa\over\dt}\int d\tau
\left[\hf\left({dx\over d\tau}\right)^2-{dx\over d\tau}\cdot\alpha\right]}
e^{i\int d\tau\left[{e\over\hbar c}\gamma^0A\bre-\beta/\lambda\right]}
\right\}\eta_i,\label{prmatr}
\ee
where $\eta_i$ and $\eta_f$ are Dirac spinors. 
The saddle point is the solution of the equation 
\be
{\delta~integrand~of~eq.~(9)\over\delta x(\tau-\dt)}=0
\ee
what can be written as
\be
{x(\tau)+x(\tau-2\dt)-2x(\tau-\dt)\over\dt^2}=
{\eta^\dagger_fF(\tau)G_0(x,y;\tau)\eta_i\over\eta^\dagger_fG_0(x,y;\tau)\eta_i},
\label{saddle}
\ee
where
\bea
F(\tau)&=&{1\over\dt}U(\tau)\biggl[L(\tau)\cdot\alpha
+{\dt\over2}\partial\gamma^0A\br
+S(\tau)T(\tau)\cdot\alpha U(\tau)\\
&&-e^{i\dt(\gamma^0A\bre-\beta/\lambda)}\biggl(L(\tau-\dt)\cdot\alpha
-{\dt\over2}U(\tau-\dt)\gamma^0\partial A\br U^\dagger(\tau-\dt)\nonu
&&+S(\tau-\dt)T(\tau-\dt)\cdot\alpha U^\dagger(\tau-\dt)\biggr)
e^{-i\dt(\gamma^0A\bre-\beta/\lambda)}\biggr]U^\dagger(\tau)\nonumber,
\eea
\bea
U(\tau)&=e^{-i\kappa\dot x(\tau)\cdot\alpha},~~~
\dot x(\tau)&={x(\tau)-x(\tau-\dt)\over\dt},\nonu
L_{jk}(\tau)&={\dot x_j(\tau)\dot x_k(\tau)\over x^2(\tau)},~~
T_{jk}&=\delta_{jk}-L_{jk},
\eea
\be
S(\tau)={\sin(\kappa|\dot x(\tau)|)\over\kappa|\dot x(\tau)|}
\ee
and $G_0$ is the integrand of \eq{prmatr} evaluated at the saddle point.
We find that the saddle point trajectory depends on the initial
and final spinors. Furthermore, the presence of the term $O(\dot x^2)$ in the
action is needed for the recovery of the usual saddle point structure.
In fact, the action $O(\dot x)$ would yield one saddle point
trajectory for each initial coordinate in contrast to the non-relativistic
limit where the initial coordinate and velocity are needed to
specify the saddle point. For a given initial coordinate $x(\tau)$
and velocity $\dot x(\tau+\dt)$ one can solve \eq{saddle} for
$x(\tau+2\dt)$. The repetition of this procedure yields the
saddle point trajectory corresponding to a given initial coordinate and velocity.
Since the integration over real trajectories with a c-number
integrand treats fermions in the same footing as bosons one can finally 
develop the same non-perturbative methods for both cases. 

This is the point where we can fix the parameter $\kappa$ which
does not influence the expectation values in the continuum limit.
It is easy to see that the straight line trajectory satisfies
\eq{saddle} for $A=0$ when $\kappa=n\pi\tau/|x-y|$. Since the average
velocity at the cutoff scale is $c$ for $\dt\to0$
the choice $\kappa=\pi$ gives the renormalized path integral which is
dominated by as smooth trajectories as possible.

Finally we mention that the term $O(\dot x^2)$ in the 
action is reminiscent of the Wilson term for lattice fermions
because it suppresses the unwanted lattice-copies. But it is given 
in the first quantized formalism and retains chiral invariance. 
Actually there is no problem of giving a 
space-time lattice regulated form of our path integral which
avoids the Nielsen-Ninomiya no-go theorem \cite{nn}. The passage over
the second quantized stage is quite a challenge. According to
the spin-statistics theorem the phase factors
corresponding to the rotation by $2\pi$ and the exchange 
of two equivalent particles are the same. Since the former
is built in our path integral formalism one expects to achieve
the correct Green functions for fermions without the use
of Grassman variables. Another interesting feature is
that the term $O(\dot x^2)$ in the action should generate
a chiral invariant Wilson term.

\end{document}